\documentclass[a4paper,11pt]{article}
\usepackage[dvips]{graphicx}
\usepackage{amssymb}
\usepackage{amsmath}
\usepackage{graphicx,,color}
\usepackage{amsfonts}
\usepackage{bm}
\usepackage{cancel}
\usepackage{comment}
\newcommand\be{\begin{equation}}
\newcommand\ee{\end{equation}}

\allowdisplaybreaks[4]

\begin{document}

\title{Novel Stellar Astrophysics from Extended  Gravity}
\author{A. V. Astashenok,$^{1}$\,\thanks{artyom.art@gmail.com}\\
$^{1}$Institute of Physics, Mathematics and IT,\\
 "I. Kant" Baltic Federal University, Kaliningrad, 236041, Russia\\
\\ \\S. Capozziello,$^{1,2}$\,\thanks{capozziello@na.infn.it}\\
$^{1}$Dipartimento di Fisica  "E. Pancini",  \\
Universit\`{a} di Napoli  "Federico II", \\
$^{2}$Istituto Nazionale di Fisica Nucleare (INFN),\\
 sez. di Napoli, Via Cinthia 9, I-80126 Napoli, Italy\\
\\ \\ S.D. Odintsov,$^{1,2}$\,\thanks{odintsov@ieec.uab.es}\\
$^{1)}$ ICREA
\\ Passeig Luis Companys
 23, 08010 Barcelona, Spain\\
$^{2)}$ Institute of Space Sciences (IEEC-CSIC)\\ C. Can Magrans
s/n,08193 Barcelona, Spain\\
\\ \\ V.K. Oikonomou$^{1}$$^{***}$\,\thanks{v.k.oikonomou1979@gmail.com,voikonomou@auth.gr}\\
$^{1)}$Department of Physics\\ Aristotle University of Thessaloniki\\ Thessaloniki 54124, Greece\\
$^{***}$Corresponding Author}

\tolerance=5000

\maketitle

%PACS numbers: 04.50.Kd, 95.36.+x, 98.80.-k, 98.80.Cq
%\pacs{04.50.Kd, 95.36.+x, 98.80.-k, 98.80.Cq,11.25.-w}

\begin{abstract}
Novel implications on neutron stars come from extended gravity. Specifically, the GW190814 event
indicated the probability of having  large mass  stars in
the mass-gap region $2.5-5\, M_{\odot}$. If the secondary
component of GW190814 is a neutron star, such large masses  are marginally supported by General Relativity (GR),
since a very stiff Equation of State (EoS) would be needed to
describe such large mass neutron stars, which would be
incompatible with the GW170817 event, without any modification of  gravity.
 In view of the two groundbreaking gravitational wave
observations, we  critically discuss the elevated role of
extensions of GR towards the successful description of the
GW190814 event, and we also speculate in a quantitative way on the
important issue of the largest allowed neutron star mass.
\end{abstract}

%%%%%%%%%%%%%%%%%%%%%%%%%%%%%%%%%%%%%%%%%%%%%%%%%%%%%%%%%%%%%%%%%%%%%%%%%%%%%%%%%%%%%%%%%%%%%%%%%%%%

Neutron stars are proven literally to be the superstars of all
stellar structures, not unjustifiably. Since the striking observation of
Jocelyn Bell back in 1967, they have been in the epicenter of many
scientific disciplines, apart from astrophysics. The last decade
two striking gravitational waves observations of the LIGO-Virgo
collaboration have set the stage for a new way of thinking both in
theoretical cosmology and in theoretical astrophysics. The
GW170817 \cite{Ligo} has proven definitive for some classes of
Horndeski theories, and also had further constrained the neutron
star masses and radii, while the GW190814 event
\cite{Abbott:2020khf} has proven even more interesting due to the
possibilities and perspectives of the secondary component
interpretation.

\begin{figure}
\centering
  \includegraphics[scale=0.35]{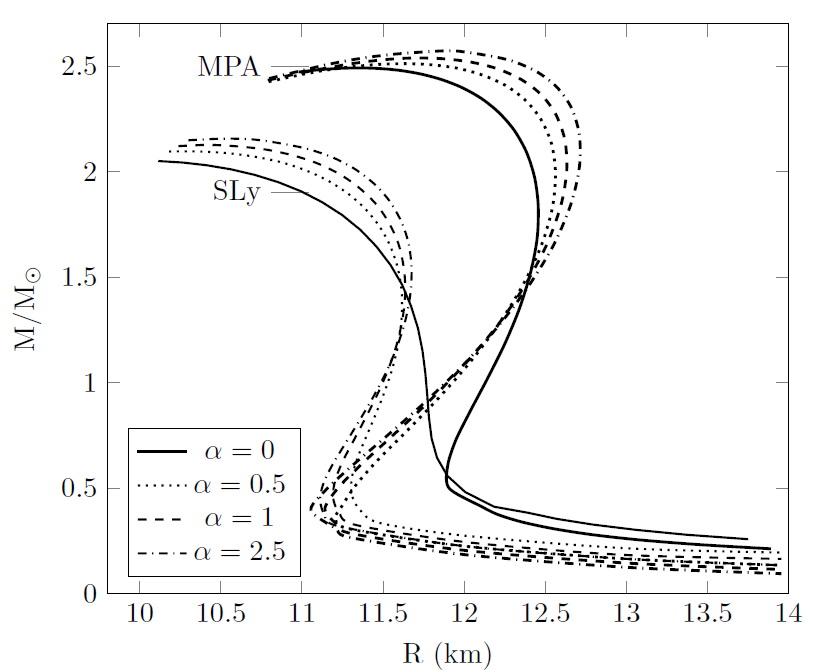}
  \caption{ $\mathcal{M-R}$ diagram for neutron stars in the context of the $R^2$ model for some standard EoS.}
  \label{fig1}
\end{figure}

Extensions of GR \cite{Report} can play a key role towards describing the
secondary component of the GW190814 event, if it is proven to be a
neutron star. The main reason is that the candidate theories
naturally yield neutron star masses in the range
$2.5-3\,M_{\odot}$, without ``stretching'' general relativistic
models. The main feature of extended gravity theories is that
geometry plays the role of the effective energy momentum tensor,
so neutron stars with masses in the range $2.5-3 \, M_{\odot}$ can
be easily described by several candidate models
\cite{Astashenok:2020qds}. In GR, where obviously modified gravity
effects are absent, one needs a quite stiff EoS for nuclear matter
to describe neutron stars with mass $2.5-3\,M_{\odot}$. On the other hand,
geometric effective contribution to the energy momentum tensor
renders modified gravity, in general, and extended gravity in particular, very appealing to describe large
mass neutron stars. Let us briefly recall the theoretical
framework of one of the most important extension of GR, namely
 $f(R)$ gravity, coinciding with the Einstein theory in the case $f(R)=R$.
 The Jordan frame gravitational action is,
\begin{equation}\label{action}
{\cal A}=\frac{c^4}{16\pi G}\int d^4x \sqrt{-g}\left[f(R) + {\cal
L}_{{\rm matter}}\right]\,,
\end{equation}
with $g$ being the determinant of the metric tensor $g_{\mu\nu}$
and ${\cal L}_{\rm matter}$  the perfect matter fluid
Lagrangian. In the context of the metric formalism, upon variation
with respect to the metric tensor, the field equations are
\begin{equation}
\frac{df(R)}{d R}R_{\mu\nu}-\frac{1}{2}f(R)
g_{\mu\nu}-\left[\nabla_{\mu} \nabla_{\nu} - g_{\mu\nu}
\Box\right]\frac{df(R)}{dR}=\frac{8\pi G}{c^{4}} T_{\mu \nu },
\label{field_eq}
\end{equation}
with $\displaystyle{T_{\mu\nu}=
\frac{-2}{\sqrt{-g}}\frac{\delta\left(\sqrt{-g}{\cal
L}_m\right)}{\delta g^{\mu\nu}}}$ being the energy momentum tensor
of the perfect  fluid matter. A spherically symmetric, static
neutron star is described by the metric
\begin{equation}
    ds^2= e^{2\psi}c^2 dt^2 -e^{2\lambda}dr^2 -r^2 (d\theta^2 +\sin^2\theta
    d\phi^2)\, .
    \label{metric}
\end{equation}
Inside the neutron star, the perfect matter fluid energy momentum
tensor is $T_{\mu\nu}=\mbox{diag}(e^{2\psi}\rho c^{2},
e^{2\lambda}p, r^2 p, r^{2}p\sin^{2}\theta)$ where $\rho$ and $p$
are the matter energy density and the pressure. The Tolman-Oppenheimer-Volkoff (TOV) equations for $f(R)$
gravity are,
\begin{equation}\label{hydro}
    \frac{dp}{dr}=-(\rho
    +p)\frac{d\psi}{dr}\,.
\end{equation}
\begin{eqnarray}
\label{dlambda_dr} \frac{d\lambda}{dr}&=&\frac{ e^{2 \lambda
}[r^2(16 \pi \rho + f(R))-f'(R)(r^2 R+2)]+2R_{r}^2 f'''(R)r^2+2r
f''(R)[r R_{r,r} +2R_{r}]+2 f'(R)}{2r \left[2 f'(R)+r R_{r}
f''(R)\right]},
\end{eqnarray}
\begin{eqnarray}\label{psi1}
\frac{d\psi}{dr}&=&\frac{ e^{2 \lambda }[r^2(16 \pi p -f(R))+
f'(R)(r^2 R+2)]-2(2rf''(R)R_{r}+ f'(R))}{2r \left[2 f'(R)+r R_{r}
f''(R)\right]}\, ,
\end{eqnarray}
\begin{equation}\label{TOVR}
\frac{d^2R}{dr^2}=R_{r}\left(\lambda_{r}+\frac{1}{r}\right)+\frac{f'(R)}{f''(R)}\left[\frac{1}{r}\left(3\psi_{r}-\lambda_{r}+\frac{2}{r}\right)-
e^{2 \lambda }\left(\frac{R}{2} + \frac{2}{r^2}\right)\right]-
\frac{R_{r}^2f'''(R)}{f''(R)}\, .
\end{equation}
One of the most important $f(R)$ gravity model is
\begin{equation}
f(R)=R+\alpha R^2, \label{fr_form_quadratic}
\end{equation}
where $\alpha$ denotes the coupling parameter
constrained by inflationary dynamics. The numerical integration
of  TOV equations for the $R^2$ model \cite{Astashenok:2020qds}
is presented in Fig. \ref{fig1} where, for larger
values of the coupling parameter $\alpha$, higher neutron
star masses result. Clearly they can naturally fill up the mass-gap region.
Thus, from a phenomenological point of view, extended gravity can
predict neutron stars with masses in the lower edge of the
mass-gap region.

Coming to the predictions of extended gravity, the fundamental
question is: How large can a neutron star mass  be?
This question is of fundamental importance, and the only consistent
way to see what is the maximum neutron star in the context of any
theory is to examine the problem by choosing the stiffest possible
EoS for  nuclear matter. The stiffest EoS,
which simultaneously respects the high density stability condition for
nuclear matter  ($\frac{dP}{d\rho}>0$) and the
subluminality condition for the speed of sound
($\frac{dP}{d\rho}\leq c^2$), is the causal EoS with the
following form,
\begin{equation}\label{causallimiteos}
P_{sn}(\rho)=P_{u}(\rho_u)+(\rho-\rho_u)c^2\, ,
\end{equation}
with $\rho_u$ being the maximum density for nuclear matter, and
$P_u(\rho_u)$  the corresponding pressure. By assuming this EoS,
 one obtains the maximum upper mass for neutron
stars, which, in the context of GR for slowly rotating neutron
stars, is \cite{Rhoades:1974fn,Kalogera:1996ci},
\begin{equation}\label{causalupperbound}
M_{max}^{CL}=3M_{\odot}\sqrt{\frac{5\times
10^{14}g/cm^{3}}{\rho_u}}\, ,
\end{equation}
thus in the context of GR, the maximum mass limit of slowly
rotating neutron stars is,
\begin{equation}\label{3solarmasslimit}
M_{max}\leq 3M_{\odot}\, .
\end{equation}
If rotation is taken into account, the causal limit of the maximum
neutron mass becomes,
\begin{equation}\label{causalrot}
M^{CL,rot}_{max}=3.89M_{\odot}\sqrt{\frac{5\times
10^{14}g/cm^{3}}{\rho_u}}\, .
\end{equation}
Therefore, in the context of GR, one expects to find neutron stars
in, but not deeply in the mass gap region $M\sim 2.5-5\,
M_{\odot}$, and specifically in the region $M\sim 2.5-3\,
M_{\odot}$, and this under extreme conditions. With regard to any
modified gravity, the question is whether the 3 solar masses upper
limit of GR is respected. Remarkably for $f(R)$ gravity, the
answer is yes \cite{Astashenok:2021peo}, as we now evince.
Consider that the nuclear matter has the following causal EoS,
\begin{equation}\label{causallimiteosnew}
P_{sn}(\rho)=P_{u}(\rho_u)+(\rho-\rho_u)v_s^2\, ,
\end{equation}
where $v_s$ is the sound speed, we shall assume that it  varies
in the range $c^2/3\leq v_s^2\leq c^2$ \cite{Sotani:2017pfj}. Also
the transition density will be assumed to be that of the SLy EoS
at $\rho_u=2\rho_0$, where $\rho_0$ is the nuclear matter
saturation density. The results of the numerical integration of
the TOV equations for the $R^2$ model are presented in Table
\ref{table1}
\begin{table}[htbp]%[H]
\begin{centering}
\begin{tabular}{|c|c|c|c|c|c|}
  \hline
  EoS  & $\alpha$,         & $M_{max}$,    & $R_{s}$, & $\Delta M_{max},$  \\
       & $r_{g}^2$    & $M_{\odot}$   & km     &  $M_{\odot}$          \\
    \hline
    \multicolumn{5}{|c|}{$\rho_u=2\rho_0$}\\\hline
                    & 0  & 1.92 & 11.28  & 0      \\
                    & 0.25 & 1.97 & 11.45 & 0.05 \\
       SLy+(5)      & 2.50  & 2.04 & 11.54  & 0.12   \\
 with $v_s^2=c^2/3$ & 10  & 2.11 & 11.69 & 0.19    \\
        \hline
                    & 0  & 2.97 & 12.85  & 0      \\
                    & 0.25 & 2.93 & 13.28 & -0.04    \\
    SLy+(5)          & 2.50  & 2.98 & 13.57  & 0.01   \\
  with $v_s^2=c^2$ & 10  & 3.10 & 13.71 & 0.13    \\
  \hline
\end{tabular}
\caption{Causal maximum mass of neutron star for the $R^2$ model.
The parameter $\alpha$ is $r_g^2=4G^2M^2_\odot/c^4$ units with
$r_g$ being the gravitational radius of the Sun.} \label{table1}
\end{centering}
\end{table}
The results of the numerical analysis are particularly
interesting. Firstly, in all studied case, the 3 solar masses
limit of GR is well respected. Let us discuss in brief the
outcomes of our analysis, starting from the effect of the
parameter $\alpha$. As it can be seen in Table \ref{table1}, for
values of the sound speed less than the speed of light, the causal
maximum mass for the $R^2$ model is larger than that of GR, while
when the sound speed is equal to the light speed, when small
values of $\alpha$ are used, the GR causal maximum mass limit is
larger compared to the $R^2$ model. On the contrary, for large
values of $\alpha$, the $R^2$ model dominates over GR. We have to
note here that the parameter $\alpha$ for the $R^2$ model in
cosmological contexts, must take small values. Specifically,  the
parameter $M$ must be approximately $1/\sqrt{\alpha}= 1.5\times
10^{-5}\left(\frac{N}{50}\right)^{-1}M_p$ for early time
phenomenological reasons \cite{Appleby:2009uf}, with $N$ being the
$e$-foldings number. So basically $\alpha$ must take small values
in order for the theory to be cosmologically consistent. This
issue has been addressed more thoroughly in Refs.
\cite{Odintsov:2021nqa,Odintsov:2021qbq} in the Einstein frame for
potentials that belong in the same class as the $R^2$ model when
considered in the Einstein frame. So we refer the reader to Refs.
\cite{Odintsov:2021nqa,Odintsov:2021qbq} for a detailed discussion
on these issues.

Thus, coming to the  issue on
where to expect to find the maximum neutron star masses, the
prediction for extended gravity is in the range $M\sim 2.5-3\, M_{\odot}$. The main
difference with GR, is the freedom in the choice of EoS for nuclear matter.

Let us note that it is remarkable that observationally-friendly
increase of neutron stars maximum mass due to extended gravity is
in compliance with inflationary attractors Universe, which is
consistent with Planck data.

Having discovered the mass range of neutron stars even in extended
gravity, the great question is: What is the lowest mass limit of
an astrophysical black hole? Without being very strict, the answer
is that astrophysical black holes can have masses above 3 solar
masses. But in order to be sure on this, one must calculate the
baryon masses for neutron stars in the context of extended
gravity. This will definitely answer, in a concrete way, the above
question, because, if the baryon mass is larger than the maximum
baryon mass, the neutron star will collapse to a black hole. Thus
automatically, one has available the lowest astrophysical black
hole mass limit. Finally, let us comment that the same procedure
performed in this paper, can be applied in other modified gravity
theories, such as $f(T)$ gravity, in which the spherically
symmetric solutions have also extended terms similar to the ones
of $f(R)$ gravity, see for example
\cite{Ruggiero:2016iaq,Ren:2021uqb}.

\end{document}